\documentclass[useAMS,usenatbib,usegraphicx,a4]{mn2e}
\usepackage{graphicx}
\usepackage{color}
\usepackage{amssymb}

\title[The Multi-periodic Blazhko modulation of CZ~Lac]{The Multi-periodic Blazhko Modulation of CZ~Lacertae\thanks{Based on observations collected with the automatic 60-cm telescope of Konkoly Observatory, Budapest, Sv\'abhegy}}

\author[\'A. S\'odor et al.]{\'A. S\'odor$^{1}$\thanks{E-mail: sodor@konkoly.hu}, J. Jurcsik$^{1}$, B. Szeidl$^{1}$, M. V\'aradi$^{2}$, A. Henden$^{3}$, K. Vida$^{1}$,\and Zs. Hurta$^{4}$, K. Posztob\'anyi$^{4}$, I. D\'ek\'any$^{1,5}$ and A. Szing$^{4}$\\
\\
$^{1}$Konkoly Observatory of the Hungarian Academy of Sciences, P.O.~Box~67, H-1525 Budapest, Hungary\\
$^{2}$Observatoire de Gen\`eve, Universit\'e de Gen\`eve, CH-1290, Sauverny, Switzerland\\
$^{3}$American Association of Variable Star Observers, 49 Bay State Road, Cambridge, MA 02138, USA\\
$^{4}$Visiting astronomer at the Konkoly Observatory of the Hungarian Academy of Sciences\\
$^{5}$Departamento de Astronom\'{\i}a y Astrof\'{\i}sica, Pontificia Universidad Cat\'olica de Chile,\\
\ \ Av. Vicu\~na Mackenna 4860, Casilla 306, Santiago 22, Chile.\\
}

\begin{document}

\date{Accepted 2010 ..... Received 2010 ...}

\pagerange{\pageref{firstpage}--\pageref{lastpage}} \pubyear{200x}

\maketitle

\label{firstpage}

%max. 200 szo
\begin{abstract}
Thorough analysis of the multicolour CCD observations of the RRab-type variable, CZ~Lacertae is presented. The observations were carried out in two consecutive observing seasons in 2004 and 2005 within the framework of the Konkoly Blazhko Survey of bright, northern, short-period RRab variables. The $O-C$ variation of CZ Lac indicated that a significant period decrease took place just around the time of the CCD observations. Our data gave a unique opportunity to study the related changes in the pulsation and modulation properties of a Blazhko star in detail. Two different period components ($\approx14.6$~d and $\approx18.6$~d) of the Blazhko modulation were identified. Both modulation components had similar strength. The periods and amplitudes of the modulations changed significantly from the first season to the next, meanwhile, the mean pulsation amplitude slightly decreased. The modulation frequencies were in a 5:4 resonance ratio in the first observing season then the frequencies shifted in opposite directions, and their ratio was close to the 4:3 resonance in the next season. The interaction of the two modulations caused beating with a period of 74~d in the first season, which resembled the 4-yr-long cycle of the $\approx40$-d modulation of RR Lyr. The mean values of the global physical parameters and their changes with Blazhko phase of both modulation components were determined by the Inverse Photometric Method.
\end{abstract}


\section{Introduction}

RR Lyrae variables have great astrophysical importance for they are tracers of stellar evolution, they are used as distance indicators and they are test objects of stellar pulsation theories. We think that these objects are known fairly well. They are pure radial pulsators oscillating in fundamental mode, in first overtone and, in some cases, in both.

There is, however, a flaw in this nice picture: the so-called Blazhko effect \citep{blazhko}, the periodic variation of the light-curve shape of some of the RR Lyrae stars. Although the phenomenon has been known for more than a century, its physical origin is still unknown. Recently, it has been found that a much larger percentage of the RR Lyrae stars exhibit light-curve modulation than it was previously assumed \citep{kbs,benko}, which makes the problem even more embarrassing.

To explain the light-curve modulation, several models have been put forward during the past decades, but none of them is able to reproduce most of the observed properties of the modulation. The oblique magnetic rotator models \citep{cousens,shiba} are based on the presumption of a bipolar magnetic field with a magnetic axis oblique to the axis of rotation. The resonance models \citep{ndz,dzm} explain the modulation with the excitation of non-radial modes of low radial degree. According to the model proposed by \cite{stothers,stothers2010}, a turbu\-lent dynamo in the envelope triggers the light-curve modulation. The analysis of the 127-d-long quasi-continuous $Kepler$ data raised the possibility that a 9:2 radial-overtone -- fundamental-mode resonance might be connected to the Blazhko phenomenon \citep{szabo}.

To find the physical explanation for the Blazhko effect, we have to know how it manifests itself in the observed properties, i.e. detailed observations of Blazhko stars are needed. Therefore, we launched the Konkoly Blazhko Survey in 2004 \citep{sodor,kbs}, which aimed at collecting extended, accurate, multicolour photometric data of Blazhko stars. In this paper, we report the results on CZ~Lacertae, one of the most intriguing target of our survey.

CZ~Lac ($m_V \approx 11.\!\!^{\rm m}7$, $\alpha_{2000} = 22^{\rm h}19^{\rm m}30.\!\!^{\rm s}76$, $\delta_{2000} = +51{\degr}28'14.\!\!^{\prime\prime}8$) is a bright, short-period, fundamental-mode RR~Lyrae-type (RRab) variable in the northern sky discovered by \cite{florja}. No photometric study of CZ~Lac has been published yet.

\section{The data}

\begin{table}
  \centering
  \caption{Log of the CCD observations of CZ~Lac obtained with the 60-cm automatic telescope. Note that, due to technical problems, observations were not obtained in each band on 10 nights in the second observing season. \label{tbl:obslog}}
  \begin{tabular}{ccclcc}
    \hline
    Season & From [JD] & To [JD] & Filter & Nights & Points \\
    \hline
    1st    & 2\,453\,266 & 2\,453\,411 & \ \ $B$            & 68 & 3669 \\
           &             &             & \ \ $V$            & 68 & 3683 \\
           &             &             & \ \ $R_\mathrm{C}$ & 68 & 3649 \\
\smallskip &             &             & \ \ $I_\mathrm{C}$ & 68 & 3629 \\
    2nd    & 2\,453\,648 & 2\,453\,731 & \ \ $B$            & 42 & 3011 \\
           &             &             & \ \ $V$            & 52 & 4460 \\
           &             &             & \ \ $R_\mathrm{C}$ & 50 & 3446 \\
           &             &             & \ \ $I_\mathrm{C}$ & 50 & 3371 \\
    \hline
  \end{tabular}
\end{table}

\subsection{Observations}

CZ~Lac has an optical companion at 10-arcsec separation, which makes low-resolution photometric observations defective.

CZ~Lac was observed  with a photometer attached to the 60-cm telescope of the Konkoly Observatory, Sv\'abhegy, Budapest, through $B$ and $V$ filters on seven nights in 1967. These observations showed light-curve variability, however, no secondary period were found. The light-curve changes were attributed to the light contamination of the close companion at that time. The photoelectric data are available online as Supporting Information (Table~S1). This table lists the differential $B$ and $V$ magnitudes of CZ~Lac relative to the  comparison star, BD~$+50{\degr} 3664$.

We observed CZ~Lac with the same, refurbished and automated 60-cm telescope equipped with a Wright Instruments $750\times1100$ CCD camera (field of view $17\times24$~arcmin) in 2004 and 2005. About 29\,000 frames were obtained in $BV(RI)_\mathrm{C}$ bands. The data of the two seasons spanned 146 and 84 d. The gap between the consecutive seasons' observations was 237-d-long. Due to problems with the filter changer, $B$, $R_\mathrm{C}$ and $I_\mathrm{C}$ observations were not obtained on each of the nights in the second season. The log of the CCD observations is presented in Table~\ref{tbl:obslog}.

To tie the instrumental CCD magnitudes to the standard Johnson--Cousins system, observations of stars in a \hbox{$20\times20$~arcmin} field centred on CZ~Lac were made with the USNO Flagstaff Station 1.0-m telescope equipped with a SITe/Tektronix  $1024\times1024$ CCD. The positions and the $BV(RI)_{\rm C}$ magnitudes of these stars are available online as Supporting Information (Table~S2).

\subsection{CCD reduction, photometry}

Standard CCD calibration was done using the program package {\sc iraf}\footnote{{\sc iraf} is distributed by the National Optical Astronomy Observatory, which is operated by the Association of Universities for Research in Astronomy, Inc., under cooperative agreement with the National Science Foundation.}. In order to eliminate the light contamination of the close companion, Image Subtraction Method (ISM), as implemented by \cite{alard}, was applied on the calibrated frames.

The relative fluxes of the variable, i.e. the differences between the measured fluxes and the flux of the variable in a reference image, were determined for each frame by aperture photometry using the  {\sc daophot} package of {\sc iraf}. To convert the relative fluxes to differential magnitudes, the fluxes and the corresponding magnitudes of the variable were measured in the reference image by aperture photometry using the standard magnitudes of 3 surrounding field stars (\hbox{USNO-A2.0 1350-16354156}, ...16333516 and ...16335838, see Table~S2). The reference magnitudes corrected for atmospheric extinction and transformed into the standard Johnson--Cousins system are listed in Table~\ref{tbl:refmag}. The standard relative $BV(RI)_\mathrm{C}$ time series of CZ~Lac are given in electronic form as Supporting Information (Tables~S3--S6). The differential magnitudes correspond to the reference magnitudes given in Table~\ref{tbl:refmag}.  

\begin{table}
  \centering
  \caption{The standard reference magnitudes of CZ~Lac corrected for atmospheric extinction in the reference images and their estimated uncertainties are given in the table.  \label{tbl:refmag}}
  \begin{tabular}{lc}
    \hline
    Filter & magnitude \\
    \hline
    \ \ $B$            & $11.83\pm0.02$ \\
    \ \ $V$            & $11.50\pm0.01$ \\
    \ \ $R_\mathrm{C}$ & $11.25\pm0.01$ \\
    \ \ $I_\mathrm{C}$ & $10.96\pm0.03$ \\
    \hline
  \end{tabular}
\end{table}

\section{Pulsation-period variations: The $O-C$ diagram}
\label{sect:oc}

\begin{table}
  \centering
  \caption{Maximum times of the Konkoly photoelectric and CCD observations.
    \label{tbl:max}}
  \begin{tabular}{ll}
    \hline
    HJD & observation \\
    \hline
    2439740.437 & photoelectric \\
    2439754.272 & photoelectric \\
    2439758.587 & photoelectric \\
    2439762.478 & photoelectric \\
    2439765.502 & photoelectric \\
    2439766.360 & photoelectric \\
    2453338.647 & CCD 2004 normal max. \\
    2453689.567 & CCD 2005 normal max. \\
    \hline
  \end{tabular}
\end{table}

Normal maximum times were derived for the middle of the observations of the two seasons' CCD $V$ band data by fitting mean pulsation light curves. These normal maximum times and the times of the 6 photoelectric $V$ maxima observed in 1967 are listed in Table~\ref{tbl:max}.

\begin{figure}
 \begin{center}
  \includegraphics[width=70mm]{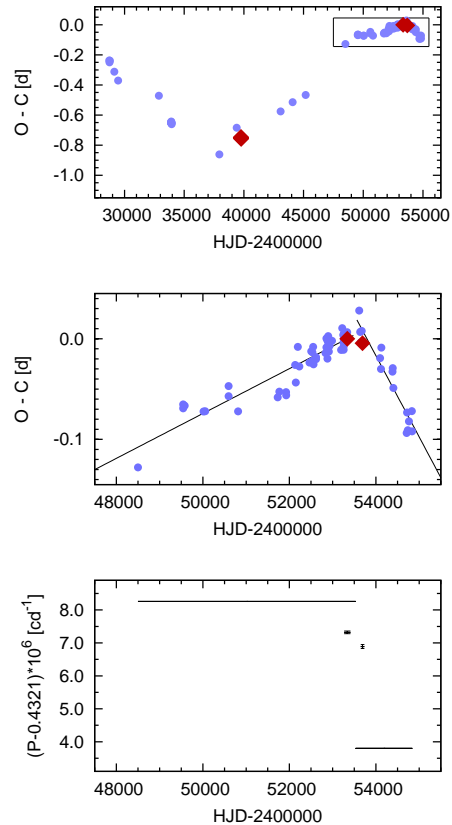}
 \end{center}
 \caption{$O-C$ diagram of CZ~Lac between 1937 and 2009. Data from the literature (collected in the GEOS data base) are plotted with dots, Konkoly observations are plotted with diamonds. The data after HJD 2\,448\,000 (1990) are shown magnified in the middle panel. These data indicate a sudden period decrease just around the time of the CCD observations (2004--2005). The straight lines fitted to the data shown in this panel prior to and after 2005.5 (HJD\,2\,453\,520) are also drawn. Bottom panel shows the period values determined from the $O-C$ data and from the Konkoly CCD observations. This figure proves that, contrary to the $O-C$ results, the pulsation period of CZ Lac did not change abruptly, but the period change lasted for several years.  \label{fig:oc}}
\end{figure}

There are 87 light maxima of CZ~Lac collected in the GEOS\footnote{http://rr-lyr.ast.obs-mip.fr/dbrr/dbrr-V1.0\_0.php} data base (see references therein). We omitted 4 discrepant, visual observations and complemented the data set with the 8 Konkoly data. The $O-C$ diagram, constructed from these maximum times, is plotted in the top panel of Fig~\ref{fig:oc}. The following ephemerida were used for the calculation:
$$ \mathrm{T}^\mathrm{(calc)}_\mathrm{max}\,\mathrm{[HJD]} = 2\,453\,338.647 + 0\fd432173\, E.$$
\noindent The initial epoch and the period were taken from the light-curve solution of the first season's CCD observations. There might be uncertainties in the epoch counts when large gaps occurred between the observations, for example, around HJD\,2\,447\,000, when observations were missing for a 3300-d-long interval.

The $O-C$ diagram shows that the pulsation period of CZ~Lac underwent a significant decrease just about when the Konkoly CCD observations were obtained. The $O-C$ plot after 1990 is shown magnified in the middle panel of Fig~\ref{fig:oc}. These data are fitted linearly before and after 2005.5 (HJD\,2\,453\,520); the corresponding pulsation periods are 0.432183 and 0.432138~d. The pulsation-period changes can be followed in the bottom panel of Fig.~\ref{fig:oc}. Periods determined from the $O-C$ data prior to and after 2005.5 and found in the two seasons' CCD observations are shown. Our CCD data indicates that a small period decrease occurred between the two observing runs (see Table~\ref{tbl:basefreq}). As the periods determined for 2004--2005 fell between the two period values obtained from the $O-C$ data, the pulsation-period change of CZ~Lac should not have bee have abrupt. The period-change event between the extrema should have lasted for several years.
 
The pulsation period of Blazhko stars often shows abrupt, irregular changes \citep{geos,m5oc}. Evidences have also been given that changes in the modulation properties are connected to the pulsation-period variations \citep[e.g., in][]{rrg2}. However, no continuous, extended observation of any Blazhko star have ever been obtained coincidentally with a rapid period change event. CZ~Lac was observed in 2004--2005, just around the time when a rapid change in its period began. It gave us a serendipitous opportunity to study such an event in detail.

\section{Light-curve analysis}

\subsection{Light-curve solutions}
\label{sect:2seasons}

\begin{figure*}
 \includegraphics[width=125mm]{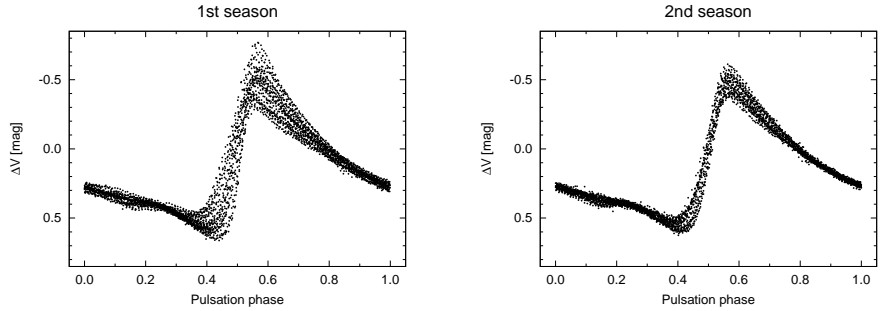}
 \caption{CCD $V$ light curves of CZ~Lac folded with the pulsation period for the two observing seasons. The properties of the modulation changed significantly between the two seasons. The overall strength of both the amplitude and phase modulations decreased.\label{fig:lcfold}}
\end{figure*}

The CCD $V$ light curves of the two observing seasons, folded with the pulsation period, are shown in Fig.~\ref{fig:lcfold}. The plots indicate that the modulation properties changed notably between the two seasons. The overall strength of the amplitude and phase modulations decreased,  which did not allow us to analyse all the data together. Therefore, the two seasons were treated separately.

A mathematical description of the brightness variations was sought in the form of Fourier sums, utilising discrete Fourier transformation and non-linear and linear least-squares fitting methods. We used the program package {\sc mufran} \citep{mufran}, the fitting abilities of {\sc gnuplot}\footnote{http://gnuplot.info/} and an own-developed code, which fitted the data with independent frequencies and their linear combinations non-linearly\footnote{http://konkoly.hu/staff/sodor/nlfit.html}.

The Fourier components were identified in the data successively until the residual spectra showed peaks higher than 4$\sigma$ in at least two colour bands at the same frequency. In addition, peaks at linear-combination frequencies of independent frequencies found in the spectra were taken into account down to about the 2$\sigma$ significance level, if they appeared in at least two bands of the same season's data.

Different components at different frequencies were needed to accurately fit the two seasons' data, but the same components were used for the $B$, $V$, $R_C$ and $I_C$ data sets in one season. The solutions of the first and second seasons' data comprised 134 and 119 frequencies, respectively.

The analysis revealed modulation components located symmetrically around the pulsation frequency ($f_\mathrm{p}$) and its harmonics in the Fourier amplitude spectra, which is characteristic of Blazhko stars. Side frequencies corresponding to modulations of two different frequencies ($f_\mathrm{m1}$ and $f_\mathrm{m2}$) were detected in both seasons. All the identified Fourier components were derived from the three independent (base) frequencies: $f_\mathrm{p}, f_\mathrm{m1}$ and $f_\mathrm{m2}$. Note that the `m1' and `m2' indices are arbitrary, and do not reflect the dominance of the $f_\mathrm{m1}$ modulation component.

Several modulation peaks appeared in the spectra around the pulsation peaks with separations of the two modulation frequencies and their linear combinations. The modulation frequencies $f_\mathrm{m1}$ and $f_\mathrm{m2}$ were also identified in the spectra. The most intriguing feature is the detection of the frequencies with separations of half of $f_\mathrm{m2}$ ($kf_\mathrm{p} \pm 0.5 f_\mathrm{m2}$). Linear combination terms involving this half frequency also appeared in the first seasons' data.

\begin{table*}
  \centering
  \caption{Independent (base) frequencies of the light-curve solutions ($f_\mathrm{p}$ -- pulsation frequency; $f_\mathrm{m1}$, $f_\mathrm{m2}$ -- modulation frequencies), their errors and their absolute and relative changes between the two seasons. The last two columns list the corresponding period values. 
\label{tbl:basefreq}}
  \begin{tabular}{cr@{.}lr@{.}lr@{.}lr@{.}lr@{.}lr@{.}l}
    \hline
                        & \multicolumn{2}{c}{1st season}     & \multicolumn{2}{c}{2nd season}     & \multicolumn{4}{c}{Freq. change}  & \multicolumn{2}{c}{1st season}     & \multicolumn{2}{c}{2nd season} \\
component               & \multicolumn{2}{c}{freq. [cd$^{-1}$]} & \multicolumn{2}{c}{freq. [cd$^{-1}$]} & \multicolumn{2}{c}{absolute [cd$^{-1}$]} & \multicolumn{2}{c}{relative} & \multicolumn{2}{c}{Period [d]} & \multicolumn{2}{c}{Period [d]} \\
    \hline
$f_\mathrm{p}$          & 2&313887  & 2&313910 &  2&$3\cdot 10^{-5}$ & 9&$9\cdot 10^{-6}$ &  0&4321732 &  0&4321689\\
\smallskip\ \ \ \ $\pm$ & 0&000002  & 0&000003 &  0&$5\cdot 10^{-5}$ & 2&$2\cdot 10^{-6}$ &  0&0000004 &  0&0000006\\
$f_\mathrm{m1}$         & 0&05406   & 0&05347  & -5&$9\cdot 10^{-4}$ &-1&$1\cdot 10^{-2}$ & 18&50      & 18&70\\
\smallskip\ \ \ \ $\pm$ & 0&00002   & 0&00007  &  0&$9\cdot 10^{-4}$ & 0&$2\cdot 10^{-2}$ &  0&01      &  0&02\\
$f_\mathrm{m2}$         & 0&06757   & 0&06954  & 19&$7\cdot 10^{-4}$ & 2&$9\cdot 10^{-2}$ & 14&804     & 14&38\\
\smallskip\ \ \ \ $\pm$ & 0&00002   & 0&00005  &  0&$7\cdot 10^{-4}$ & 0&$1\cdot 10^{-2}$ &  0&004     &  0&01\\
    \hline
  \end{tabular}
\end{table*}

\begin{table*}
  \centering
  \caption{Frequencies, amplitudes (in mag) and phases (in rad) of the $B, V, R_\mathrm{C}$ and $I_\mathrm{C}$ light-curve solutions of CZ~Lac in the first observing season. The full table is available as Supporting Information with the online version of this article. \label{tbl:2004solution}}
  \begin{tabular}{clllllllll}
    \hline
    \multicolumn{2}{c}{ }    & \multicolumn{2}{c}{$B$} & \multicolumn{2}{c}{$V$} & \multicolumn{2}{c}{$R_\mathrm{C}$} & \multicolumn{2}{c}{$I_\mathrm{C}$} \\
    identification         & $f$ [cd$^{-1}$] & amplitude & phase    & amplitude & phase       & amplitude & phase    & amplitude & phase       \\
    \hline
    $1f_\mathrm{p}$               & 2.313887 & 0.4894(6) & 3.393(1) & 0.3535(5) & 3.330(1)    & 0.2763(4) & 3.248(2) & 0.2110(5) & 3.117(2) \\
    $2f_\mathrm{p}$               & 4.627774 & 0.2537(6) & 2.909(2) & 0.1895(5) & 2.896(2)    & 0.1500(5) & 2.868(3) & 0.1137(5) & 2.822(4) \\
    $3f_\mathrm{p}$               & 6.941660 & 0.1472(6) & 2.737(4) & 0.1115(5) & 2.731(4)    & 0.0891(5) & 2.726(5) & 0.0691(6) & 2.728(7) \\
    ... \\
    $f_\mathrm{m1}$               & 0.054065 & 0.0056(4) & 1.51(7)  & 0.0035(3) & 1.40(10)    & 0.0035(3) & 1.50(9)  & 0.0035(3) & 1.84(9) \\
    $f_\mathrm{m2}/2$             & 0.033786 & 0.0025(4) & 0.8(2)   & 0.0024(3) & 0.5(1)      & 0.0029(3) & 0.9(1)   & 0.0040(3) & 0.87(9) \\
    $f_\mathrm{m2}$               & 0.067573 & 0.0041(4) & 3.0(1)   & 0.0013(3) & 2.5(2)      & 0.0027(3) & 2.3(1)   & 0.0023(4) & 1.4(1)  \\
    $1f_\mathrm{p}-f_\mathrm{m1}$ & 2.259822 & 0.0235(5) & 1.34(2)  & 0.0172(4) & 1.33(2)     & 0.0137(4) & 1.41(3)  & 0.0107(4) & 1.52(4) \\
    $2f_\mathrm{p}-f_\mathrm{m1}$ & 4.573709 & 0.0272(5) & 0.67(2)  & 0.0202(4) & 0.69(2)     & 0.0159(4) & 0.74(2)  & 0.0122(4) & 0.78(3) \\
    ... \\
    $1f_\mathrm{p}+f_\mathrm{m1}$ & 2.367951 & 0.0515(5) & 3.36(1)  & 0.0378(4) & 3.41(1)     & 0.0289(4) & 3.46(1)  & 0.0223(4) & 3.53(2) \\
    $2f_\mathrm{p}+f_\mathrm{m1}$ & 4.681838 & 0.0448(5) & 2.73(1)  & 0.0331(4) & 2.77(1)     & 0.0270(4) & 2.78(2)  & 0.0208(5) & 2.79(2) \\
    ... \\
    $1f_\mathrm{p}-f_\mathrm{m2}$ & 2.246314 & 0.0136(4) & 0.85(3)  & 0.0105(3) & 0.84(3)     & 0.0091(3) & 0.83(4)  & 0.0071(4) & 0.79(5) \\
    $2f_\mathrm{p}-f_\mathrm{m2}$ & 4.560201 & 0.0160(4) & 0.25(3)  & 0.0117(3) & 0.33(3)     & 0.0102(3) & 0.37(3)  & 0.0080(4) & 0.35(4) \\
    ... \\
    $1f_\mathrm{p}+f_\mathrm{m2}$ & 2.381459 & 0.0391(4) & 4.99(1)  & 0.0291(3) & 5.03(1)     & 0.0229(3) & 5.00(1)  & 0.0165(4) & 4.96(2) \\
    $2f_\mathrm{p}+f_\mathrm{m2}$ & 4.695346 & 0.0316(4) & 4.37(1)  & 0.0238(4) & 4.39(1)     & 0.0179(3) & 4.42(2)  & 0.0144(4) & 4.54(3) \\
    ... \\
    \hline
  \end{tabular}
\end{table*}

\begin{table*}
  \centering
  \caption{Frequencies, amplitudes (in mag) and phases (in rad) of the $B,V, R_\mathrm{C}$ and $I_\mathrm{C}$ light-curve solutions of CZ~Lac in the second observing season. The full table is available as Supporting Information with the online version of this article.\label{tbl:2005solution}}
  \begin{tabular}{clllllllll}
    \hline
    \multicolumn{2}{c}{}    & \multicolumn{2}{c}{$B$} & \multicolumn{2}{c}{$V$} & \multicolumn{2}{c}{$R_\mathrm{C}$} & \multicolumn{2}{c}{$I_\mathrm{C}$} \\
    identification         & $f$ [cd$^{-1}$] & amplitude & phase    & amplitude & phase       & amplitude & phase    & amplitude & phase       \\
    \hline
    $1f_\mathrm{p}$       &     2.313910   & 0.4854(5) & 2.863(1) & 0.3509(4) & 2.802(1) & 0.2749(4) & 2.721(1) & 0.2046(4) & 2.577(2) \\
    $2f_\mathrm{p}$       &     4.627821  & 0.2495(5) & 1.843(2) & 0.1863(4) & 1.830(2) & 0.1469(4) & 1.808(3) & 0.1094(4) & 1.783(4) \\
    $3f_\mathrm{p}$       &     6.941731  & 0.1454(5) & 1.131(4) & 0.1109(4) & 1.121(3) & 0.0886(4) & 1.132(4) & 0.0689(4) & 1.139(6) \\
    ... \\
    $f_\mathrm{m1}$       &  0.053469   & 0.0011(4) & 4.0(4)   & 0.0015(3) & 4.1(2)   & 0.0015(3) & 4.0(2)   & 0.0014(3) & 4.2(2)   \\
    $f_\mathrm{m2}$      &   0.069541    & 0.0036(4) & 2.9(1)   & 0.0018(3) & 3.7(2)   & 0.0018(3) & 4.7(2)   & 0.0016(3) & 4.4(2)   \\
    $1f_\mathrm{p}-f_\mathrm{m1}$& 2.260441 & 0.0088(6) & 4.17(8)  & 0.0055(4) & 4.59(9)  & 0.0039(4) & 4.3(1)   & 0.0033(5) & 4.3(2)   \\
    $2f_\mathrm{p}-f_\mathrm{m1}$& 4.574351 & 0.0094(5) & 3.15(6)  & 0.0062(4) & 3.25(7)  & 0.0041(4) & 3.2(1)   & 0.0032(4) & 3.5(1)   \\
    ... \\
    $1f_\mathrm{p}+f_\mathrm{m1}$& 2.367380 & 0.0191(6) & 5.33(3)  & 0.0151(4) & 5.19(3)  & 0.0118(4) & 5.24(4)  & 0.0094(5) & 5.42(5)  \\
    $2f_\mathrm{p}+f_\mathrm{m1}$& 4.681290 & 0.0137(5) & 4.26(4)  & 0.0120(4) & 4.18(3)  & 0.0107(4) & 4.27(4)  & 0.0082(5) & 4.41(6)  \\
    ... \\
    $1f_\mathrm{p}-f_\mathrm{m2}$&  2.244370 & 0.0079(6) & 1.14(7)  & 0.0060(4) & 0.85(6)  & 0.0051(4) & 0.86(7)  & 0.0045(5) & 1.1(1)   \\
    $2f_\mathrm{p}-f_\mathrm{m2}$&  4.558280 & 0.0088(6) & 6.23(7)  & 0.0072(4) & 6.08(5)  & 0.0061(4) & 6.03(6)  & 0.0049(4) & 6.1(1)   \\
    ... \\
    $1f_\mathrm{p}+f_\mathrm{m2}$& 2.383451 & 0.0242(6) & 3.80(3)  & 0.0183(4) & 3.88(2)  & 0.0142(4) & 3.85(3)  & 0.0109(4) & 3.73(4)  \\
    $2f_\mathrm{p}+f_\mathrm{m2}$& 4.697362 & 0.0241(6) & 2.56(3)  & 0.0176(4) & 2.62(2)  & 0.0136(4) & 2.61(3)  & 0.0102(4) & 2.57(4)  \\
    ... \\
    \hline
  \end{tabular}
\end{table*}

\begin{figure*}
 \begin{center}
  \includegraphics[height=100mm, bb=50 50 607 377]{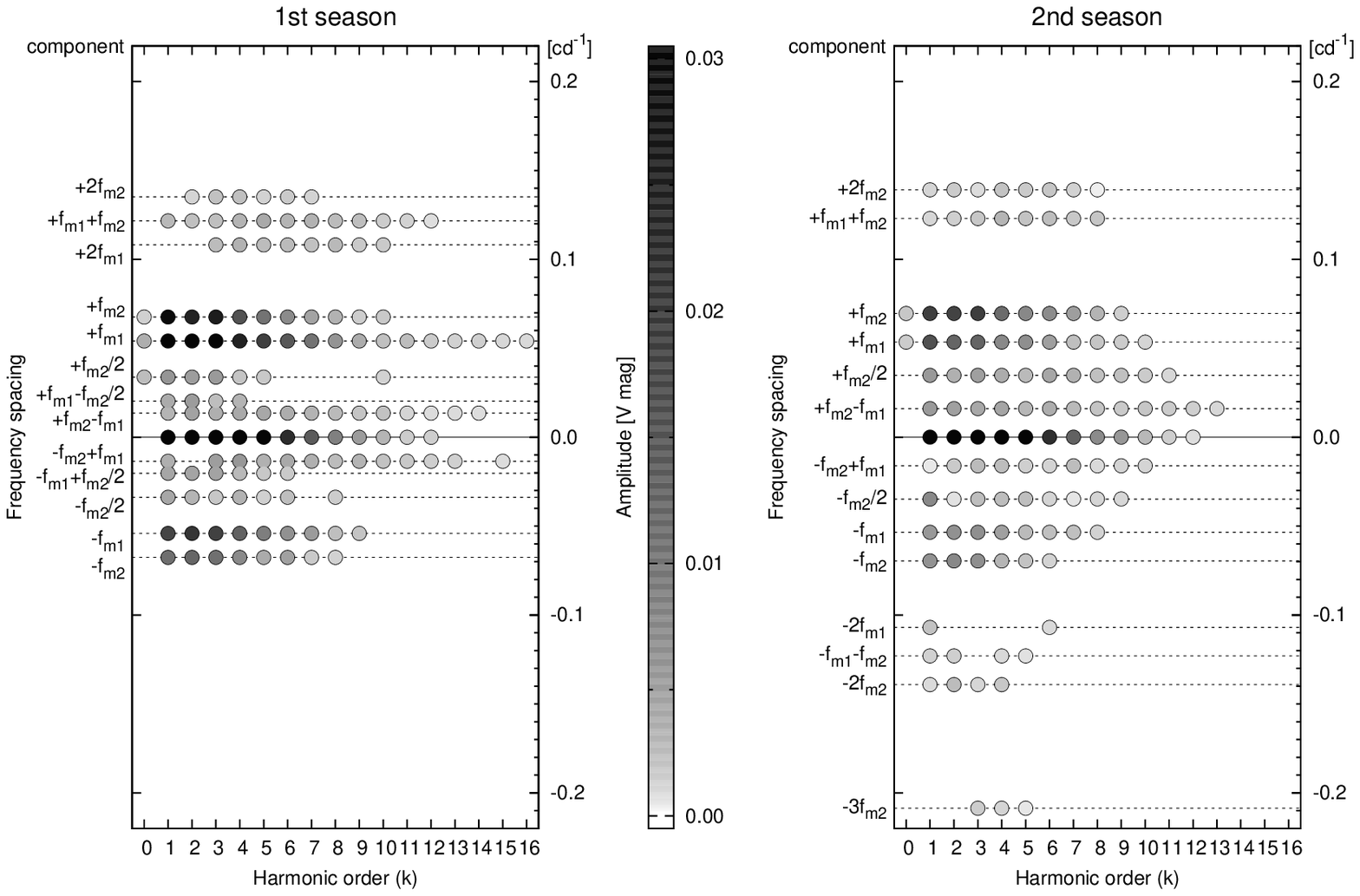}
  \caption{Multiplet structures of the light-curve solutions of CZ~Lac in the two observing seasons. Horizontal axes show the pulsation harmonic order ($k$). The vertical axes show the vicinity of the pulsation harmonic frequencies. The identification of the frequency components are given. The $V$ amplitudes are colour-coded; amplitudes higher than 0.03~mag are shown full black. These figures demonstrate that significantly different frequency components are needed to fit the two seasons' data. Note the discernible changes in the amplitudes and separations of the modulation components.
  \label{fig:lincombs}}
 \end{center}
\end{figure*}

Having  the relevant Fourier components identified, their frequency values were fixed to the values obtained from the $V$-band light curve solutions as the signal-to-noise properties were the best in this band. The $V$-band solutions were obtained by non-linear fits that allowed to adjust the three independent base frequencies, and the combination frequencies were locked to the base-frequency values during the fitting process. The base frequencies, their absolute and relative changes and the formal 1$\sigma$ errors of the $V$-band solutions are given in Table~\ref{tbl:basefreq} for the two seasons.

To determine the amplitudes and phases of the frequency components in the four bands, linear fits were calculated with all the frequencies fixed to their accepted values. The parameters of the solutions are listed in Tables~\ref{tbl:2004solution}~and~\ref{tbl:2005solution}. The formal $1\sigma$ errors of the Fourier amplitudes and phases are also given.

The detected frequencies are illustrated schematically in Fig.~\ref{fig:lincombs}. The complexity of the multiplet structures in the Fourier spectra, and the changes in the frequencies, amplitudes and in the appearing linear combination terms between the two seasons can be well followed in this figure.

The residual spectra showed no signal stronger than 4$\sigma$, except one peak in each of the $B$, $R_\mathrm{C}$ and $I_\mathrm{C}$ bands, which appeared, however, at different frequencies in the first season's data. These peaks were most probably artefacts and were not intrinsic to the variable. Inspecting the residual light curves folded with the pulsation period, we found extended scatter around the phase of the middle of the rising branch. Similar behaviour of the residuals of other Blazhko stars were shown in \cite{mw1} and in \cite{poretti}. 

The Fourier frequency components $kf_\mathrm{p} \pm 0.5 f_\mathrm{m2}$ are extraordinary. No similar peak has ever been detected in any other Blazhko star. A recent study on the long-term behaviour of Blazhko stars in the globular cluster, M5 \citep{m5bl} revealed that V2/M5 had a modulation period of 67~d at around the middle of the 20th century, while  its Blazhko cycle turned to be 130~d, just about twice of its previous value, after 1960. Anomalous modulation components with $11.5 f_\mathrm{m}, 12.5 f_\mathrm{m}$ and $13.5 f_\mathrm{m}$ separations were also detected in MW~Lyr \citep{mw1}. These examples indicate that the half of the modulation frequency might have some relevance in the interpretation of the data. It cannot even be excluded that $0.5 f_\mathrm{m2}$ is one of the real modulation frequencies of CZ Lac and peaks with separations of $f_\mathrm{m2}$, $2f_\mathrm{m2}$ and $3f_\mathrm{m2}$ are, in fact, its 2nd, 4th and 6th modulation harmonics. The side-lobe peaks with $0.5 f_\mathrm{m2}$ separations around the pulsation harmonics are, however, much weaker than those with $f_\mathrm{m2}$ separations. Therefore, keeping in mind the possibility that the true modulation frequency might be $0.5 f_\mathrm{m2}$, for practical purposes we have chosen the large-amplitude component,  $f_\mathrm{m2}$, as the first harmonic-order component of this modulation.

\subsection{Changes between the seasons}
\label{sect:changes}

\begin{figure*}
  \includegraphics[height=54mm]{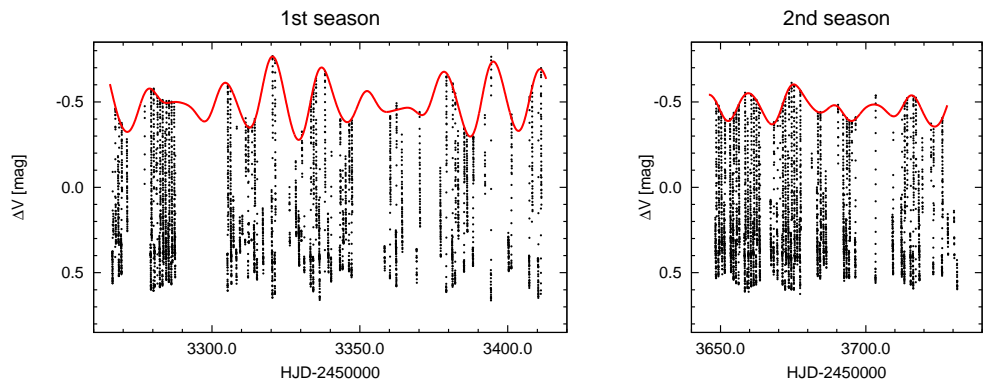}
 \caption{CCD $V$ light curves of CZ~Lac versus HJD. The upper envelope of the data shows the brightness variations of the light maxima. The properties of the modulation changed considerably between the two seasons. The overall strength of the modulation decreased.\label{fig:lcnonfold}}
\end{figure*}

 It was shown in the previous sections that a slight increase of the pulsation period ($0.0000043 \pm 0.0000010$~d) was accompanied by significant frequency and amplitude changes of the complex modulation of CZ~Lac. The CCD $V$ magnitudes of the two seasons are plotted against HJD  in Fig.~\ref{fig:lcnonfold}. The upper envelopes derived from the synthetic light curves are also drawn. The interaction between the two modulations produced beating in the first season; they  cancelled  each other out periodically as can be seen around HJD\,2\,453\,290 and HJD\,2\,453\,365. The envelope of the second season shows a weaker amplitude modulation.

The $kf_\mathrm{p} \pm f_\mathrm{m1}$ modulation components had somewhat higher amplitudes than the amplitudes of the $kf_\mathrm{p} \pm f_\mathrm{m2}$ frequencies in the first season. The amplitudes of both modulations decreased to the second season by different rates, so that the amplitude relation of the two modulations reversed. The amplitudes of the $kf_\mathrm{p} \pm f_\mathrm{m2}$ modulation components were larger than those of the $kf_\mathrm{p} \pm f_\mathrm{m1}$ components then.

It seemed that no changes occurred in the frequencies and amplitudes within one season, so the pulsation and modulations of CZ~Lac could be described reasonably well by stationary harmonic functions in both seasons. The first season's data were extended enough to be divided into two parts at HJD\,2\,453\,350. A separate analysis indicated that no systematic changes in the frequencies and/or amplitudes occurred in the first observing season. Had the amplitudes and frequencies changed somewhat during any of the observing seasons, the residual spectra would show significant peaks around the subtracted frequencies, which was not the case. These arguments support that the changes took place dominantly between the two seasons, within a relatively short interval of 237 d.

Not only the amplitudes of the modulation components but also the amplitude of the pulsation changed between the two seasons. The peak-to-peak mean pulsation amplitudes of both seasons were determined from the synthetic light curves taking into account only the pulsation harmonics. Their errors were estimated as the sum of the errors of the amplitudes of the pulsation harmonic components. Because the errors of the amplitudes were correlated and the phases were not taken into account, these errors overestimated the true uncertainties. We found that the mean pulsation amplitude decreased by $0.027\pm0.011$, $0.023\pm0.009$, $0.015\pm0.008$ and $0.017\pm0.009$\,mag in $B$, $V$, $R_\mathrm{C}$ and $I_\mathrm{C}$ bands, respectively, simultaneously with the diminution of the modulation amplitudes. The epoch-independent phase differences of the pulsation harmonics  did not change significantly between the two seasons, indicating that the shape of the mean pulsation light curve remained stable, whilst its amplitude decreased.

Simultaneous pulsation-period and -amplitude changes were also detected in RR~Gem around the year 1938 \citep{rrg2}, when the character of the modulation of the star changed as well. The pulsation period and amplitude increased in RR~Gem, while they decreased in CZ~Lac.

\subsection{Resonances; one or two modulation frequencies?}

The frequency ratio of the two modulations was $f_\mathrm{m2}/f_\mathrm{m1} = 1.2499\pm 0.0008$ in the first season, i.e. the frequencies were in a 5:4 resonance ratio well within the error limit. As Table~\ref{tbl:basefreq} shows, the modulation frequencies changed in opposite directions between the two seasons, and as a consequence escaped from the 5:4 resonance. The frequency ratio of the two modulations was $1.301 \pm 0.002$ in the second season, which was closer to the 4:3 resonance ratio than to the 5:4 one.

The transition from the 5:4 resonance to the near 4:3 resonance hints that the two modulation frequencies prefer to be in resonance. Considering this finding, one might wonder if indeed two linearly independent modulation frequencies exist in CZ Lac. Could the light curves be described with only one modulation frequency and many of its harmonics, as, for example, in MW~Lyr \citep{mw1} and in V1127~Aql \citep{v1127aql}? In this case, the base modulation frequencies of the two seasons were significantly different, they were $f_\mathrm{m}=0.006757\,\mathrm{cd^{-1}} = f_\mathrm{m1}/8 = f_\mathrm{m2}/10$ and $f_\mathrm{m}=0.008802\,\mathrm{cd^{-1}}\approx f_\mathrm{m1}/6\approx f_\mathrm{m2}/8$ in the first and second seasons, respectively. The residual scatters of the one-modulation-frequency fits are the same in the first season, and they are only 3--5~per~cent larger in the second season than those of the two-modulation-frequency solutions. The one-modulation solution would also give a natural explanation for the appearance of the combination terms. However, serious problems emerge when one tries to interpret this solution.

One of the problems is that it is ambiguous how to relate the different modulation components of the two seasons' solutions in this case. The one-modulation-frequency solutions incorporate different-order modulation sidelobe frequencies ($kf_\mathrm{p}\pm nf_\mathrm{m}$) in the two seasons. Components corresponding to $n=\{2, 3, 5, 8, 10, 16, 18, 20\}$ and $n=\{2, 4, 6, 8, 12, 14, 16, 24\}$ are involved in the first and second seasons' solutions, respectively. In contrast, if we accept the two-modulation-frequency solution, a consistent identification of the modulation components can be set, and the connection between the frequencies of the two seasons is straightforward.

Another problem of the one-modulation-frequency solution is that, in this case, the modulation harmonics of CZ~Lac would behave rather differently than those in other Blazhko stars. In well-observed Blazhko stars with quintuplets and higher-order multiplet structures detected,  the frequency components at the smallest separation ($n=1$) had always the largest amplitude (for example, V1127~Aql -- \citealt{v1127aql}, DM~Cyg -- \citealt{dmc}, MW~Lyr -- \citealt{mw1}, RV~UMa -- \citealt{rvuma}). In contrast, according to the one-modulation solution of CZ~Lac the largest-amplitude, dominant modulation peaks are the $n=8$ and 10 components in the first season and the $n=8$ and 6 components in the second one. The $n=1$ component is not even detectable in any of the seasons. It has been shown in \citet{coast} that the $f_\mathrm{p}\pm f_\mathrm{m}$ frequencies are always the largest-amplitude modulation components in the frequency spectrum of an amplitude- and phase-modulated harmonic oscillation, provided that the amplitude of the phase modulation is relatively small compared to the period of the oscillation. Note that the same argument applies against the acceptance of $0.5 f_\mathrm{m2}$ as the base modulation frequency of the second modulation.

All these arguments lead us to the conclusion that indeed two independent modulations of CZ~Lac coexisted and their frequencies changed in opposite directions between the two observing seasons.
 
\begin{table*}
\centering
\caption{Mean physical parameters of CZ~Lac and their errors derived from the mean light curves by the inverse photometric method, using static atmosphere models of \hbox{[Fe/H] $= -0.20$} metallicity.}
\label{tbl:averagephispar}
\begin{tabular}{ccccccccc}
\hline
${M}$           & $d$        & $E(B-V)$        & $M_V$         & $(B-V)_0$       & $(V-I)_0$       & $R$           & $L$          & $T_\mathrm{eff}$ \\ 
{[M$_\odot$]} & [pc]       & [mag]           & [mag]         &  [mag]          &   [mag]         & [$R_\odot$]   & [$L_\odot$]  &        [K]       \\
\hline
$0.60\pm0.01$            & $1188\pm8$ & $0.19 \pm 0.03$ & $0.66\pm0.01$ & $0.276\pm0.002$ & $0.344\pm0.002$ & $4.46\pm0.03$ & $46.5\pm0.5$ & $7100\pm10$      \\
\hline
\end{tabular}
\end{table*}

\begin{figure*}
 \includegraphics[width=150mm,height=84mm]{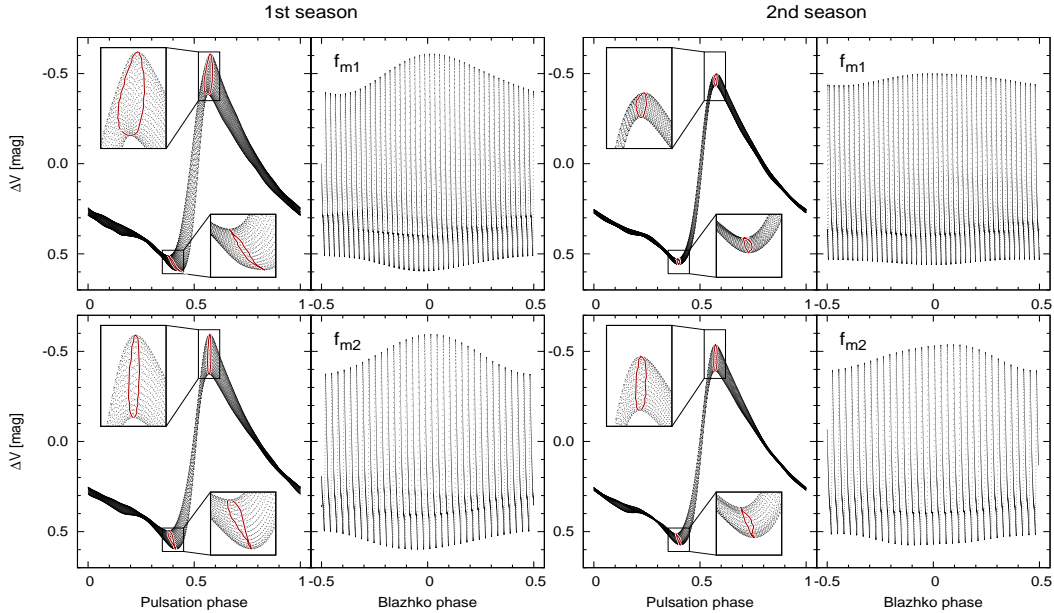}
 \caption{Synthetic $V$ light curves of CZ~Lac corresponding only one of the modulation components are phased according to the pulsation and modulation frequencies for the two seasons. These data were used as input for the IPM to determine the variations of the pulsation-averaged physical parameters with Blazhko phase. Lines indicate the phase and brightness variations in maximum and minimum light.\label{fig:slcs}}
\end{figure*}

\section{Resemblance between the modulations of CZ~Lac and RR~Lyr}

The $\approx40$-d-long Blazhko modulation of RR~Lyr  seemed to be ceased in every 4~years \citep{dsz73}. The latest weak-modulation phase had been detected in 1975 \citep{sz76} in the Konkoly observations, which continued until 1981. Although RR Lyr was not regularly observed in the past three decades, photometric observations were obtained in 1990--1993 \citep{sz97}, 1993--1994 \citep{bels97}, 1996 \citep{smith03}, 2003--2004 \citep{kk1} and 2006--2007 \citep {kk2}. All these observations showed maximum-light variation in $V$ band of at least 0.2-mag amplitude during the Blazhko cycle. The amplitude of the modulation was also large in 2009 according to the recent $Kepler$ observations \citep{szabo}. The spectroscopic observations from 1994--1995 \citep{cg96}, 1994--1997 \citep{ch00}, 1996--1997 \citep{ckag99} and 2000 \citep{cc06}, all showed clear evidence of strong amplitude modulation. It seems that, in spite of the above mentioned observational activities, the weak-modulation states of RR~Lyr have been either missed or the cyclic behaviour of the modulation has been less pronounced during the last decades.

Variation in the modulation period of RR~Lyr was also reported by \cite{kk1,kk2}. In contrast with the \hbox{40--41\,d} Blazhko period found in previous investigations, a \hbox{38--39\,d} period was detected in these recent studies.

Many aspects of the Blazhko behaviour of CZ~Lac resembles the properties of RR~Lyr. The modulation of CZ~Lac  seemed to cease in every 74~days in the first season, as a consequence of the interaction between two, similarly strong modulations. This beating is shown by the envelope in Fig.~\ref{fig:lcnonfold}. The beating period is $1/(f_\mathrm{m2}-f_\mathrm{m1})=74.02$~d. We suppose that the four-year cycle of RR~Lyr can also be explained by the beating of two modulations of similar strength, if the periods of the modulations differ by about 1~d. To resolve modulation peaks being so close to each other, continuous observations of more than four years are needed, which will be available from the $Kepler$ mission.

The strength of the modulation components of CZ~Lac changed significantly from the first observing season to the next; while the 18.5-d modulation ($f_\mathrm{m1}$)  had somewhat larger amplitude than the 14.8-d modulation ($f_\mathrm{m2}$) in the first season, the amplitude of the longer-period component ($f_\mathrm{m1}$) was extremely small in the second one. Consequently, no cancellation (beating) of the modulations was detected then. A similar behaviour might explain the vanishing of the 4-yr cycles of RR~Lyr during the last decades and the strong change in its observed  modulation period.

\section{Changes in physical parameters}
We have developed
an inverse photometric Baade--Wesselink method (IPM) \citep{ip} to determine the mean physical parameters of RR Lyrae stars, such as effective temperature ($T_\mathrm{eff}$), absolute visual magnitude ($M_\mathrm{V_0}$), radius ($R$), distance ($d$), mass (${M}$) and dereddened colours [$(B-V)_0$, $(V-I_\mathrm{C})_0$], from multicolour photometric observations. It has been shown that using good quality multicolour photometry, IPM yields results of similar accuracy as direct spectroscopic Baade--Wesselink methods. The IPM also gives the variations with pulsation phase of those parameters that change during the pulsation cycle. This method has been successfully applied to derive the mean physical parameters and their changes during the Blazhko cycle already for four modulated RRab stars (MW~Lyr -- \citealt{mw2}, DM~Cyg -- \citealt{dmc}, RR Gem, SS Cnc -- \citealt{aipc}). The IPM is a useful tool to unravel astrophysics buried deep in the hundreds of Fourier coefficients obtained by the light-curve-fitting process.

In order to determine the changes with any of the modulations of CZ~Lac by the IPM, the two modulations had to be separated first. It was achieved by generating synthetic light curves from the solutions of each season's data taking into account only the pulsation frequencies and terms that belonged to one of the modulations only. The coupling modulation terms were omitted, as they were not periodic with any of the modulations. Considering their small amplitudes, the omission of these terms presumably had a negligible effect on the results. The synthetic $V$ light curves of the two modulations for both seasons are plotted in Fig.~\ref{fig:slcs}.

\subsection{Constant parameters}

Following the methodology of our previous investigations, we started the analysis by determining those parameters that obviously did not change. These were the metallicity ([Fe/H]), mass ($\mathfrak{M}$), distance ($d$) and the interstellar reddening [$E(B-V)$] of the star. 

The metallicity of CZ~Lac was determined spectroscopically \citep{suntzeff} and photometrically \citep{kbs} to be $-0.13$ and $-0.26$, respectively. Therefore, static atmosphere model grids corresponding to $\mathrm{[Fe/H]}=-0.20$ \citep{cast} were used for the IPM analysis.

With the exception of the metallicity, all the other constant parameters were derived by the IPM using the mean pulsation light curves, which were generated from the light-curve solutions taking into account only the pulsation components. To assess the uncertainties of the mean global physical parameters and the distance, the IPM was run with four different internal settings \citep[for details see][Table~1]{ip} using the mean light curves of the two observing seasons. These runs yielded eight results for each physical parameter. Their averages are listed in Table~\ref{tbl:averagephispar}. The errors of the output parameters, with the exception of $E(B-V)$, were estimated as the rms of the corresponding eight results. The error of the derived interstellar reddening depended on the calibration of our photometry, i.e. on the accuracy of the standard reference magnitudes given in Table~\ref{tbl:refmag}. The IPM yielded an interstellar reddening value of $E(B-V)=0.19$~mag for CZ~Lac in accordance with the 0.28\,mag value given by the reddening map of \cite{schlegel}. The eight different results for the mass and the distance of CZ~Lac were within the ranges of 0.590--0.618\,M$_\odot$ and 1182--1205\,pc, respectively. These narrow ranges proved the stability of the results obtained with the IPM.

Possible changes in the mean radius, luminosity and temperature between the two seasons were also checked. After fixing the constant parameters (metallicity, mass, distance and interstellar reddening) to their values determined in the previous step, IPM was run again for the mean pulsation light curves of the two seasons separately. The results did not show any significant variation in any of the free parameters between the two seasons. This finding was in accordance with the photometric results, which showed that the mean $BV(RI)_\mathrm{C}$ colours of the light curves remained the same within the millimagnitude limit in the two seasons. The differential magnitude zero-points of the light-curve solutions for the two seasons are compared in Table~\ref{tbl:magzp}.

\begin{table}
  \centering
  \caption{Differential magnitude zero-points of the solutions for the two observing seasons in the four photometric bands. Relative values with respect to the reference magnitudes listed in Table~\ref{tbl:refmag} are given.}
  \label{tbl:magzp}
  \begin{tabular}{ccc}
    \hline
    band & 1st season     & 2nd season \\
    \hline
    $B$            & 0.2902(2) & 0.2907(3) \\
    $V$            & 0.1538(3) & 0.1544(3) \\
    $R_\mathrm{c}$ & 0.1327(3) & 0.1346(2) \\
    $I_\mathrm{c}$ & 0.1073(3) & 0.1086(2) \\
    \hline
  \end{tabular}
\end{table}

\begin{figure*}
 \begin{centering}
  \includegraphics[width=82mm]{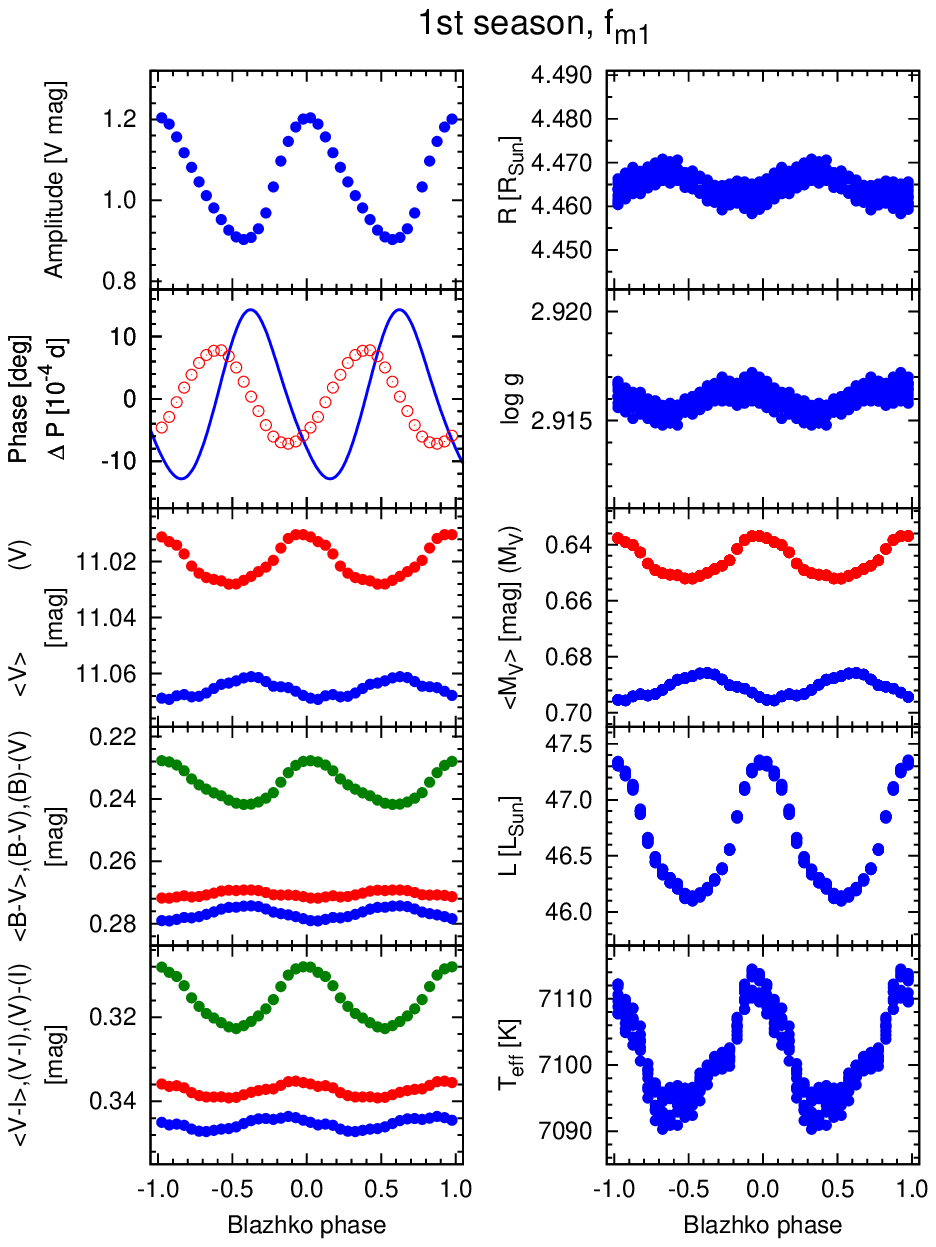}\includegraphics[width=82mm]{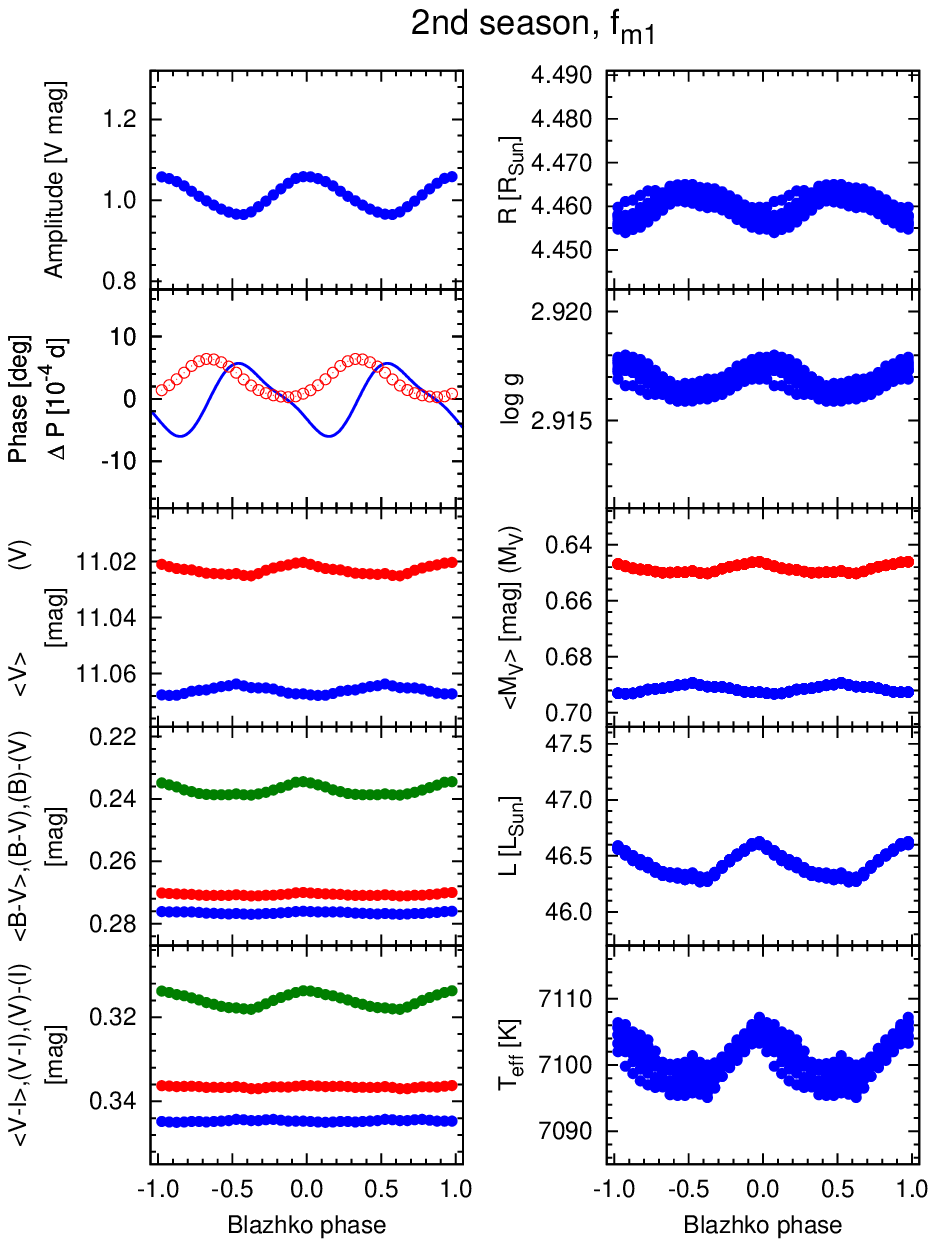}\\
  \includegraphics[width=82mm]{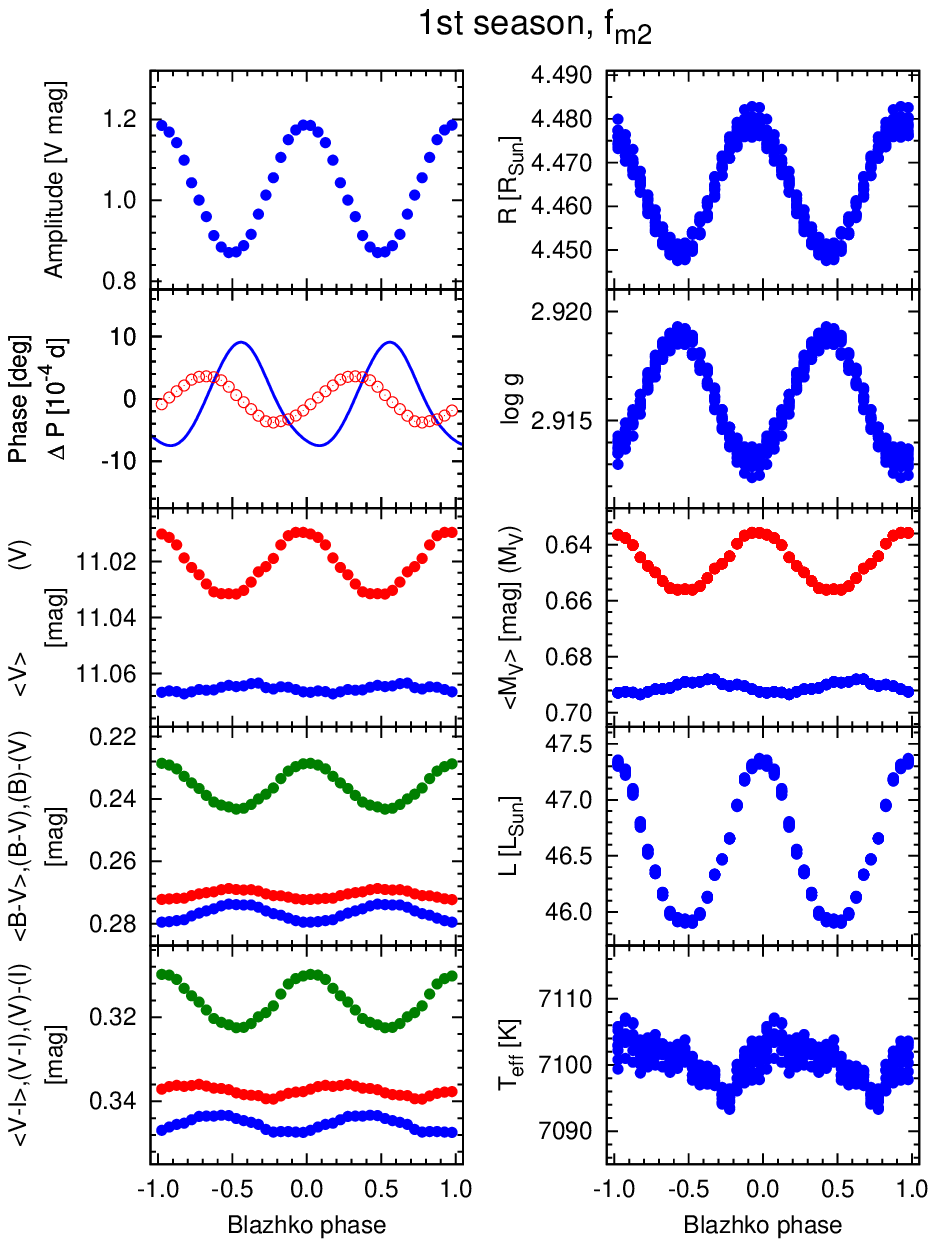}\includegraphics[width=82mm]{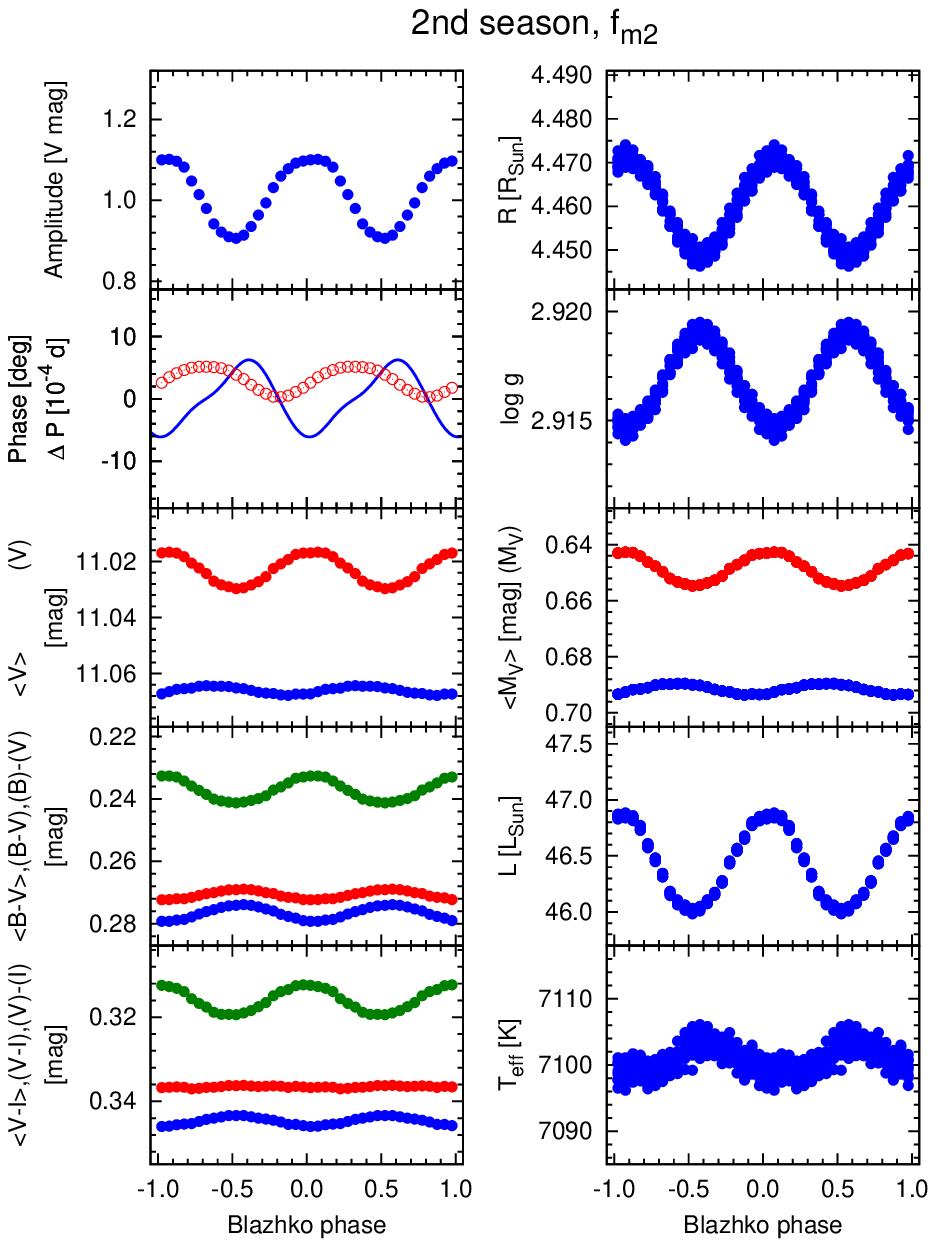}\\
  \caption{Variations of the observed mean (left-hand panels) and derived physical parameters (right-hand panels) of CZ~Lac during the Blazhko cycles. Each of the four blocks of panels corresponds to one modulation component ($f_\mathrm{m1}$ or $f_\mathrm{m2}$) in one season. Magnitude- and intensity-averaged brightnesses and colours are denoted by angle and round brackets, respectively. From top to bottom, left-hand panels show : amplitude modulation -- total pulsation amplitude; pulsation-phase or period modulation -- variation of the phase of the $f_\mathrm{p}$ pulsation component (empty circles) and deviation of the instantaneous pulsation period from the mean pulsation period (continuous line); pulsation-averaged $V$ magnitude; pulsation-averaged $B-V$ colour; pulsation-averaged $V-I_\mathrm{C}$ colour. From top to bottom, right-hand panels show the pulsation averages of the following physical parameters : radius; surface gravity; absolute visual magnitude; luminosity; effective surface temperature. \label{fig:ipres}}
 \end{centering}
\end{figure*}

\subsection{Blazhko-phase dependent parameters}

After determining the mass, the distance and the interstellar reddening of CZ~Lac, the IPM were run for the data of different phases of the modulations, keeping these parameters fixed during the fitting process. In this way, the changes in radius, luminosity and temperature with Blazhko phase were obtained for both modulations in the two observing seasons.

The synthetic light curves of both modulations were divided into 20 Blazhko-phase bins of equal width. These data showed practically no light-curve modulation in any of the bins. The light curves of the modulation phase bins were then fitted with the harmonics of the pulsation frequency. These fits were used as the input for the IPM. The results, the pulsation-phase-dependent output parameters of the IPM were averaged for the whole pulsation cycle, i. e. their arithmetic means were calculated in order to detect any Blazhko-phase-dependent changes. At this stage, eight different internal settings of the IPM were used to estimate the uncertainties of the results. The choices of the $V_\mathrm{rad}$ template curves and their weights during the fitting process were the same that we had used in our earlier analyses \citep[see][Table~1]{ip}. Furthermore, we used two different $A(V_\mathrm{rad})$--$V_\mathrm{amp}$ relations \citep[for details see][]{mw2}.

The mean values of the observational data, derived from the photometry, and the results of the IPM are plotted in Fig.~\ref{fig:ipres} for both modulations in both seasons. In the left-hand panels, the observed mean quantities are plotted as a function of Blazhko phase: the peak-to-peak pulsation amplitude in $V$ band, the variation of the pulsation period calculated from the phase variation of the $f_\mathrm{p}$ pulsation-frequency, and the different pulsation averages of the magnitudes and colours. The right-hand panels show the quantities derived by the IPM for each Blazhko phase: the pulsation-averaged mean values of the radius, the surface gravity, the absolute visual brightness averaged by magnitude and intensity units, the luminosity and the effective temperature. In full accordance with our previous findings \citep{mw2,dmc}, the actual values of the fixed parameters had an influence only on the average values of those physical parameters that changed with Blazhko phase, and had no effect on their changes around the averages.

As the amplitudes of both modulations decreased from the first season to the next, the amplitudes of the changes in most of the derived parameters decreased, too. Although the two modulations had rather different characteristics, apart from the reduced amplitudes, their behaviour remained the same from the first season to the second. The consistency of the IPM results for the two modulations in the different seasons is a strong argument for the reliability of the method. Taking into account that the input data of the IPM are completely independent for the two seasons, such a consistent result can hardly be explained by any artefact.

In accordance with our previous results \citep{mw2,dmc,aipc}, the intensity-averaged mean $V$ brightness reflects the mean luminosity changes of CZ~Lac very well. The luminosity is always the highest at the high-amplitude Blazhko phases. The luminosity variations connected to the first modulation ($f_\mathrm{m1}$) are caused predominantly by temperature changes in both seasons. The radius changes corresponding to this modulation component are very small and are in anti-correlation with the luminosity variations. At the same time, the luminosity changes connected to the second modulation ($f_\mathrm{m2}$) are induced mainly by radius changes in both seasons. The small temperature changes with this modulation are out of phase in the first season and are the opposite of the luminosity variations in the second season.

The strength of the phase (period) modulations seems to have no influence on the strength of any of the detected variations of the parameters.

\section{Discussion and summary}

Multi-periodic modulation has already been found in some Blazhko RR~Lyr stars \citep{xzcyg,uzu,asasbl,lsh,benko}. However, one of the modulations was always dominant, with much higher amplitude than the other, in these stars. CZ~Lac is the first multi-periodically modulated RR~Lyrae star with two modulations of similar strength. The good phase coverage of our multicolour CCD observations made it possible to study the nature of the multi-periodic modulation of CZ Lac in detail.

The analysis of the data from two consecutive observing seasons revealed that the Fourier spectra of the light curves showed complex multiplet structures around the pulsation harmonics; frequencies at $kf_\mathrm{p} \pm f_\mathrm{m1}$, $kf_\mathrm{p} \pm f_\mathrm{m2}$, $kf_\mathrm{p} \pm (f_\mathrm{m1}+f_\mathrm{m2})$, $kf_\mathrm{p} \pm (f_\mathrm{m1}-f_\mathrm{m2})$, $kf_\mathrm{p} \pm 2f_\mathrm{m1}$, $kf_\mathrm{p} \pm 2f_\mathrm{m2}$, $kf_\mathrm{p} - 3f_\mathrm{m2}$ were detected. One of the most remarkable properties of the multi-periodic modulation of CZ Lac was that, similarly to multi-mode pulsation, combination terms of the modulations also appeared in the spectra. Modulation peaks with separations of half the $f_\mathrm{m2}$ frequency and linear combination terms involving this component [$kf_\mathrm{p} \pm 0.5 f_\mathrm{m2}$ and $kf_\mathrm{p} \pm (f_\mathrm{m1} - 0.5 f_\mathrm{m2})$] were also identified. No similar frequency component has ever been detected in any other Blazhko star. No additional frequency at $0.5 f_\mathrm{m}$ separation has been reported even in the high-precision $CoRoT$ and $Kepler$ space missions' data \citep{v1127aql,benko}. The appearance of the $0.5 f_\mathrm{m2}$ components makes the interpretation of the spectrum rather difficult. Taking also into account that modulation periods of 130~d and about its half, 67~d were detected at different epochs in one of the RRab stars (V2) in the globular cluster M5 \citep{m5bl}, one might have some doubt whether $f_\mathrm{m2}$ or its half is the true base frequency of this modulation.

The modulation properties of CZ~Lac changed significantly between the two observing seasons, while only very slight changes were detected in the mean pulsation properties. The periods of the two modulations changed in opposite directions: $P_\mathrm{m1}$ increased from 18.5 to 18.7~d, while $P_\mathrm{m2}$ decreased form 14.8 to 14.4~d. The amplitudes of both modulations decreased significantly from the first observing season to the next, and their relative strength reversed. The period and the $V$ amplitude of the mean pulsation light curve decreased by $4\cdot 10^{-6}$~d and 0.015~mag, respectively. All the detected changes took place within a 237-d-long interval, as there was no sign of any frequency or amplitude change in the individual seasons' data.

The two modulation frequencies of CZ Lac were in a 5:4 resonance ratio well within the error limit in the first observing season. The modulation frequencies changed to the second season so that they were close to the 4:3 resonance then, but the frequency ratio differed somewhat from the exact resonance value. The sudden transition from a 5:4 resonance to a near 4:3 resonance between the two seasons hints that the two modulations prefer to be in resonance and their frequencies are not fully independent. However, the strong, opposite period changes of the components have made it impossible to describe the light-curve variations of CZ~Lac uniformly with a single base-modulation frequency in the two seasons. There is another example of a Blazhko star (V14 in M5) with modulation periods in 4:3 resonance \citep{m5bl}. However, the modulations with the different periods were not coincidental, they were observed at different epochs in this star. These observations suggest that the detected modulation periods of Blazhko stars may not be unique, which makes the interpretation of the phenomenon even more difficult.

Changes in the physical parameters of CZ~Lac were determined by the Inverse Photometric Method \citep{ip}. No changes in the mean parameters were derived for the two consecutive seasons in accordance with the constancy of the mean colours and magnitudes during the whole two-years observations. Small changes in the mean effective temperature ($T_\mathrm{eff}$), mean luminosity ($L$) and mean radius ($R$) with the phases of both modulations were unambiguously detected. The relations between the physical-parameter variations were different for the two modulations in both seasons. Although the large-amplitude phases of both $f_\mathrm{m1}$ and $f_\mathrm{m2}$ corresponded to 1--3~per~cent increased mean luminosity values, these luminosity changes were induced by either temperature or radius changes connected to one and the other modulations. This pattern remained unchanged from the first season to the next. Taking into account the independence of the input data (observations of the two seasons) and their separate analysis, such stability of the results confirms that the detected changes are real.

CZ Lac is an example for modulations of different character showing up simultaneously in a single star. Consequently, it seems that the stellar structure and parameters do not directly determine how the mean physical parameters vary with the Blazhko cycle. This fact may explain why no generally valid relation between pulsation and modulation properties of Blazhko stars has been revealed in spite of the efforts for establishing such a connection. 

The phase relation between the pulsation-period and -amplitude variations during the Blazhko cycle should be an important parameter of the modulation according to the model of \cite{stothers,stothers2010}. Examining these phase relations of the stars studied so far in the Konkoly Blazhko Survey and the data of the two modulations of CZ~Lac, we have not found any connection between these properties and any of the observed and derived (by the IPM) parameters of the modulation. The only regularity that have been disclosed is that, in all the studied cases, the changes in the mean luminosity are always parallel to the amplitude variations of the pulsation, i.e. Blazhko stars are always slightly more luminous at their large-amplitude phase than at their small-amplitude phase.

The investigation of the long-term period changes using $O-C$ data showed that the pulsation period of CZ~Lac underwent a significant decrease just around the time of our CCD observations in 2004 and 2005. Linear fits to the $O-C$ data for the 1990--2005.5 and 2005.5--2010 intervals yielded pulsation periods of 0.432183 and 0.432138~d, respectively. The CCD observations revealed that the pulsation-period change was not abrupt, but lasted for several years. The Konkoly CCD observations were made serendipitously at the time when the drastic period decrease started. Our data showed that the modulation properties (periods and amplitudes) changed  significantly in 2004--2005, when the drastic pulsation-period decrease of CZ~Lac just began. All these changes were presumably triggered by structural changes in the stellar interior. The strong variations observed in the modulation properties implied that the modulation was more sensitive to the changes in the interior and/or followed these changes more quickly than the properties of the pulsation. A similar, significant period change (an increase, however) occurred in RR~Gem \citep[][Fig.~1]{rrg2} around 1938, which was also accompanied by drastic changes in the modulation properties.

CZ~Lac is the first Blazhko RR~Lyrae star that showed two different modulations simultaneously with amplitudes of about equal strengths.
If the two modulations are indeed independent of each other, then it contradicts any explanation of the Blazhko effect that connect the modulation period to the rotation of the star.

\section*{Acknowledgments}

The support of OTKA grant T-068626 is acknowledged.

\label{lastpage}
\end{document}